\documentclass[12pt]{iopart}

\usepackage{graphicx}
\begin{document}

\title[]{Interface-engineered hole doping in Sr$_2$IrO$_4$/LaNiO$_3$ heterostructure}

\author{Fangdi Wen$^1$, Xiaoran Liu$^1$, Qinghua Zhang$^2$, M. Kareev$^1$, B. Pal$^1$, Yanwei Cao$^3$, J. W. Freeland$^4$, A. T. N'Diaye$^5$, P. Shafer$^5$, E. Arenholz$^5$, Lin Gu$^2$, J. Chakhalian$^1$}

\address{$^1$Department of Physics and Astronomy, Rutgers University, Piscataway, New Jersey 08854, USA}
\address{$^2$Beijing National Laboratory for Condensed-Matter Physics and Institute of Physics, Chinese Academy of Sciences, Beijing 100190, P. R. China}
\address{$^3$Ningbo Institute of Materials Technology and Engineering, Chinese Academy of Sciences, Ningbo, Zhejiang 315201, China}
\address{$^4$Advanced Photon Source, Argonne National Laboratory, Argonne, Illinois 60439, USA}
\address{$^5$Advanced Light Source, Lawrence Berkeley National Laboratory, Berkeley, California 94720, USA}

\ead{fw113@physics.rutgers.edu}
\ead{xiaoran.liu@rutgers.edu}
\vspace{10pt}
\begin{indented}
\item[]\today
\end{indented}

\begin{abstract}
The relativistic Mott insulator Sr$_2$IrO$_4$  driven by large spin-orbit interaction is known for the $J_\textrm{eff}$ = 1/2 antiferromagnetic state which closely resembles the electronic structure of parent compounds of superconducting cuprates. Here, we report the realization of hole-doped Sr$_2$IrO$_4$ by means of interfacial charge transfer in Sr$_2$IrO$_4$/LaNiO$_3$ heterostructures. 
 X-ray photoelectron spectroscopy on Ir 4f edge along with the  X-ray absorption spectroscopy at Ni $L_2$ edge confirmed that 5$d$ electrons from Ir sites are transferred onto Ni sites, leading to markedly electronic reconstruction at the interface.  Although the Sr$_2$IrO$_4$/LaNiO$_3$ heterostructure remains non-metallic, we reveal that the transport behavior is no longer described by the Mott variable range hopping mode, but by the Efros-Shklovskii model.  These findings highlight a powerful utility of interfaces to realize emerging  electronic states of the Ruddlesden-Popper phases of Ir-based oxides. 
\end{abstract}

\vspace{2pc}

\submitto{New Journal of Physics}

\maketitle
%
%

\section{Introduction}

Transition metal oxides (TMOs) with a partially filled d-shell often host strongly correlated carriers and exhibit unique physical properties due to the intertwined lattice, charge, orbital and spin degrees of freedom.\cite{imada1998metal} In 3$d$ TMOs, since the crystal field (CF) splitting $\Delta_\textrm{CF}$ is approximately an order of magnitude larger than the intrinsic spin-orbit coupling (SOC) $\lambda$, the effect of SOC is usually neglected in the determination of the ground state. However, in 4$d$ and 5$d$ TMOs,  the enhanced strength of $\lambda$  becomes comparable to $\Delta_\textrm{CF}$.  As the outcome of the competing interactions  dominated by SOC, a large number of unusual quantum states including topological insulators\cite{uchida2018topological}, quantum spin liquids \cite{balents2010spin}, Weyl semimetals  \cite{rau2016spin}, and Kitaev materials\cite{witczak2014correlated} have been recently predicted. In this category, Sr$_2$IrO$_4$ is one of the prototypical examples of materials  known as the relativistic Mott insulators. Naively, due to the spatially extended $5d$ orbitals and the decreased  on-site Coulomb repulsion, a metallic ground state is naturally expected in this compound. Contrary to  the expectation, Sr$_2$IrO$_4$ is antiferromagnetic insulator. \cite{chikara2009giant} To explain the discrepancy, the proposed physical picture suggests that the degeneracy of the Ir 5$d$ levels is first lifted by the CF splitting while the strong SOC further splits the t$_{2g}$ bands into  fully occupied J$_\textrm{eff}$ = 3/2 subbands and a half-filled J$_\textrm{eff}$ = 1/2 subband. Additionally, the on-site Coulomb interaction further splits the J$_\textrm{eff}$ = 1/2 band into an upper Hubbard band (UHB) and a lower Hubbard band (LHB), realizing the spin-orbit assisted Mott ground state. \cite{kim2008novel,kim2009phase}

Based on this picture, a strong similarity between Sr$_2$IrO$_4$ and the parent componds of high T$_c$  cuprates has been highlighted in various experiments. Specifically, Sr$_2$IrO$_4$ crystallizes in the K$_2$NiF$_4$-type structure and is an antiferromagnetic insulator with the magnetic transition temperature T$_c$ $\sim$ 240 K. \cite{kim2009phase,cao1998weak} Resonant inelastic x-ray scattering experiments reveales that the magnetic excitations of Sr$_2$IrO$_4$ on the square lattice can be well described within an antiferromagnetic Heisenberg model\cite{kim2012magnetic} akin to  the parent compounds of cuprates. Furthermore,  it has been proposed that upon  hole and electron doping, this  material can potentially host a high T$_c$ superconducting state.\cite{wang2011twisted,meng2014odd,watanabe2013monte}

 Recent measurements have provided many evidences for the possible superconducting hidden phase in Sr$_2$IrO$_4$. \cite{yan2015electron,cao2016hallmarks,Zhao2016Evidence,AT2015Collapse,Kim2016Observation}  Despite the early encouraging progress, to date,  there is no report on superconducting behavior in doped Sr$_2$IrO$_4$.\cite{chen2018unidirectional,korneta2010electron,JR2016Ambipolar,LM2015Tuning} 
 Naturally, one of the possible reasons  could be that the degree of doping has not reached the critical value for the superconducting state to emerge. However, there has been chemical doping reaching the level of the theory prediction (around $\sim 12 \%$ or more), but the superconducting phase is yet to be found.\cite{watanabe2013monte,cao2016hallmarks,Zhao2016Evidence,JR2016Ambipolar,LM2015Tuning} Considering that the high chemical doping level might, on the other hand, also affect many crytallographic properties of the system, searching for alternative ways to realize such a high doping  in Sr$_2$IrO$_4$ without atoms replacement might be vital to identify the hidden superconducting  phase with iridates.     


In this letter, we report on creating high hole doping of Sr$_2$IrO$_4$ by virtue of engineering a heteroepitaxial interface composed of Sr$_2$IrO$_4$ and LaNiO$_3$ ultra-thin layers. Scanning transmission electron microscopy (STEM) and X-ray reflectivity (XRR) results confirm high quality, good crystallinity, well-formed interfaces and expected periodicity of the heterostructures. Photoelectron spectroscopy data (XPS) and X-ray absorption spectra (XAS) reveal distinct charge modulations at the interface, leading to electron transfer from Ir to Ni sites across the interfaces. X-ray linear dichroism studies have revealed  that the charge redistribution lifts the orbital degeneration in Ni$^{2+}$, and therefore created the Ni$^{2+}$ ($S$ = 1) and Ir$^{5+}$ ($J_\textrm{eff}$ = 0) electronic configurations at the interface. Although the heterostructure remains insulating, we found that the transport behavior is no longer described by the Mott variable range hopping model, but by the Efros-Shklovskii model.

\section{EXPERIMENTAL SECTION}

High-quality  Sr$_2$IrO$_4$/LaNiO$_3$ superlattices were epitaxially synthesized  on (001) oriented SrTiO$_3$ substrate by pulsed laser deposition  (KrF excimer laser, $\Lambda$=248 nm) with substrate temperature of $\sim$690 $^\circ$C and oxygen pressure of 10 mTorr. Targets were ablated at the laser frequency of 2Hz and fluency of 1 - 4 J/cm$^2$. The layer-by-layer growth was monitored by in-situ high-pressure reflection high-energy election diffraction (RHEED).

\begin{figure*}
    \includegraphics[width=0.9\textwidth]{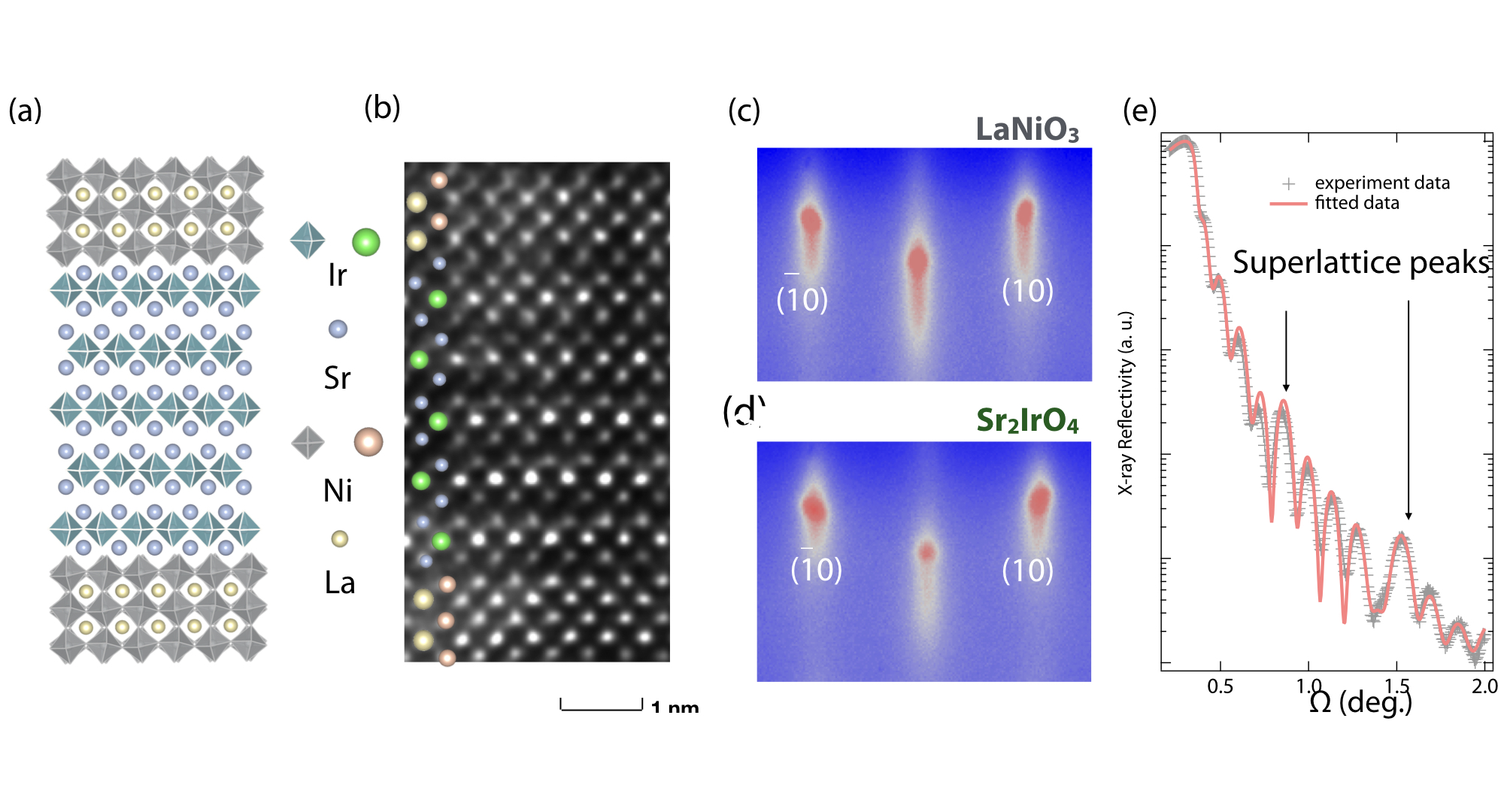}
    \caption{(a) The schematic picture of Sr$_2$IrO$_4$/LaNiO$_3$ superlattice and the interface, designated to match the TEM result shown in (b). (b) STEM image of a layer of Sr$_2$IrO$_4$ with its interface between LaNiO$_3$ upper and lower layers. Electron scattering off oxygen atoms too low to be observed.  (c) is the RHEED pattern of LaNiO$_3$ and (d) is the RHEED pattern of Sr$_2$IrO$_4$ during growth. (e) XRR data of Sr$_2$IrO$_4$/LaNiO$_3$. The superlattice peaks are indicated by black arrows, with thickness fringes visible between them.}
    \label{RHEED}
\end{figure*}

The crystallinity and epitaxy of the sample has been verified by  several methods. First, the samples have been investigated by STEM (see Fig. \ref{RHEED} (b)). As seen, within the LaNiO$_3$ layers, the La distribution shows a typical perovskite structure. In the Sr$_2$IrO$_4$, two atomic planes of SrO are observed between each layer of IrO$_2$ in according with the expected layered-perovskite structure. Sharp interfaces between Sr$_2$IrO$_4$ and LaNiO$_3$ resulting from the layer-by-layer structure of the heterostructure are also evident from the STEM images. The STEM result is in a good agreement with the in-situ RHEED measurements for both Sr$_2$IrO$_4$ and LaNiO$_3$ layers (see Fig. \ref{RHEED}(c) and (d)). Furthermore, the XRR measurement confirms the superlattice structure and allow to estimate the total film thickness, as shown in Fig. \ref{RHEED} (e). The total thickness obtained from the fitting result is $\sim$30.768 ($\pm$ 0.13) nm  while the thickness of each individual layer is $\sim$6.221 ($\pm$0.014) nm.

\begin{figure*}
    \includegraphics[width=\textwidth]{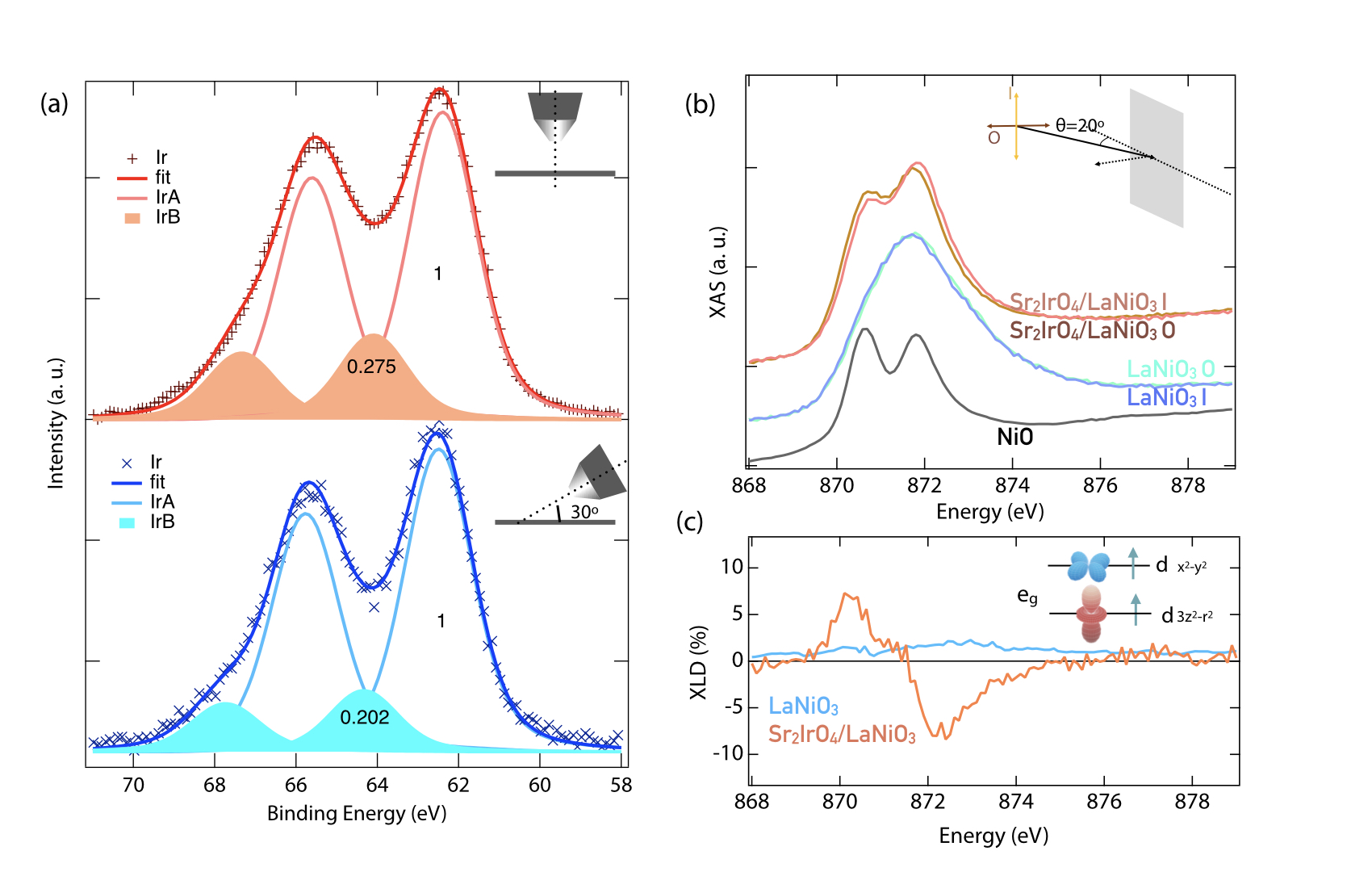}
    \caption{(a) Angle-resolved Ir 4$f_{5/2}$ and 4$f_{7/2}$ XPS of Sr$_2$IrO$_4$/LaNiO$_3$ superlattice, with the inset figure indicating different detection geometry. For comparison, the IrB doublet is normalized to the IrA doublet, which is set to unity.
    (b) Polarization-dependent XAS and (c) XLD results at Ni L$_2$ edge, together with data of pure LaNiO$_3$ films on SrTiO$_3$ substrate and NiO. 
    (c)The red line indicates XLD of Sr$_2$IrO$_4$/LaNiO$_3$ heterostructure and the blue line indicates the LaNiO$_3$ single layer. Inset is the schematic diagram of Ni $e_g$ orbitals in bulk LaNiO$_3$ and in the superlattice.}
    
    \label{XPSXLD}
\end{figure*}

Next, to understand the electronic structure associated with the Ir charge state at the Sr$_2$IrO$_4$/LaNiO$_3$ interfaces, we carried out a set of  XPS measurements with different detection angle.
Figure \ref{XPSXLD} (a) shows the Ir 4$f$ core level spectrum of the superlattice. 
From the direct fitting of the spectral shape, one can see that the Ir 4f envelope contains two doublet contributions. 
Specifically, the main doublet appearing at binding energy 62.3 eV and 65.6 eV represents the Ir$^{4+}$ states which are expected for the stoichiometric undoped system (IrA).
 There is also another doublet contribution (IrB) with binding energy close to 64 eV and 67.3 eV, which can result from the formation of Ir$^{5+}$, or accompanied satellite peaks of Ir$^{4+}$\cite{banerjee2017observationIr5+,kumar2017evolutionIr5+,zhu2014enhancedIr5+,XL2017Synthesis}. 
In order to prove the existence of Ir$^{5+}$, angle-resolved XPS  was conducted with two different detection angles. By changing the detection geometry, the XPS can show signals with different probing depth. If the IrB doublet is purely the satellite peaks, the relative intensity to the main doublet IrA should be a fixed value at different detection angle. However, there is a clear variation of IrB relative intensity as the detector is changed from normal to grazing position. This observation confirms the  presence of Ir$^{5+}$ in Sr$_2$IrO$_4$ layers at the interfacial region.

Next, we turn to investigate the charge distribution on Ni in the vicinity of the interface by XAS at the Ni L$_2$ edge. 
 As seen in a typical scan depicted in Fig. \ref{XPSXLD} (b) there is a clear double peak structure at  the L$_2$ edge in the superlattice with either horizontal- or vertical-polarized X-rays (orange curve).  This observation is in sharp contrast to the reference LaNiO$_3$ sample with pure Ni$^{3+}$, confirming the presence of Ni$^{2+}$ due to the charge transfer. 
 By fitting the XAS result of the superlattices with respect to the pure Ni$^{3+}$ and Ni$^{2+}$ reference samples, the contribution of Ni$^{2+}$ is $\sim$ 70\%, which indicates a large amount charge transfer happening at the interfaces.
 
Soft x-ray absorption with linearly polarized light or X-ray linear dichroism (XLD) allows for further insight in to the orbital  occupation of the Ni$^{2+}$.  Experimentally, the XLD is defined as the difference between absorption spectra measured with horizontally polarized light (O) where the polarization points out of the plane, and in-plane polarized light (I) with polarization in the plane, or I$_O$-I$_I$. As shown in Fig. \ref{XPSXLD} (c), in the undoped LaNiO$_3$ layer (before charge transfer)  Ni$^{3+}$ is in 3d$^7$ state with the $t_{2g}$ band fully occupied and $e_g$ band occupied with one electron. After electron transfer nickel ions shift the Ni valence state down towards Ni$^{2+}$ with two electrons occupying the e$_g$ orbitals. In our experimental setup, out-of-plane polarized X-rays probe the empty out-of-plane orbitals ($d_{3z^2-r^2}$) while in-plane polarized light senses primarily $d_{x^2-y^2}$ orbital character. The observed XLD signal can be attributed to the difference in orbital occupation of  $d_{3z^2-r^2}$ vs. $d_{x^2-y^2}$ orbitals.  From  the XAS data shown in Fig. \ref{XPSXLD} (b) and the XLD data shown in Fig. \ref{XPSXLD} (c), one can see that the absorption of photons with horizontal polarization (O) is shifted lower in energy relative to photons with vertical polarization  (I) shifted to a higher energy. This result implies that the interface imposes a sub-band splitting  of $\sim$ 0.1 eV $\bigtriangleup e_g$ which lowers the Ni $d_{3z^2-r^2}$  and lifts the $d_{x^2-y^2}$ orbital states (see Fig. \ref{XPSXLD} (c) inset). Further, 
the integrated XLD area is close to zero suggesting that both orbitals are equally occupied.\cite{cui2014tuning,pesquera2012surface,chakhalian2011asymmetric}  
In short, the XAS/XLD measurements confirms the electrons transferred from Ir to Ni sites.  This leads to (i) formation of Ni$^{2+}$ with high spin $S$=1 configuration at the interface, and (ii) split of the Ni $e_g$ band into two sub-bands  with predominant $d_{x^2-y^2}$ orbital and $d_{3z^2-r^2}$ orbital character.

\begin{figure*}
    \includegraphics[width=\textwidth]{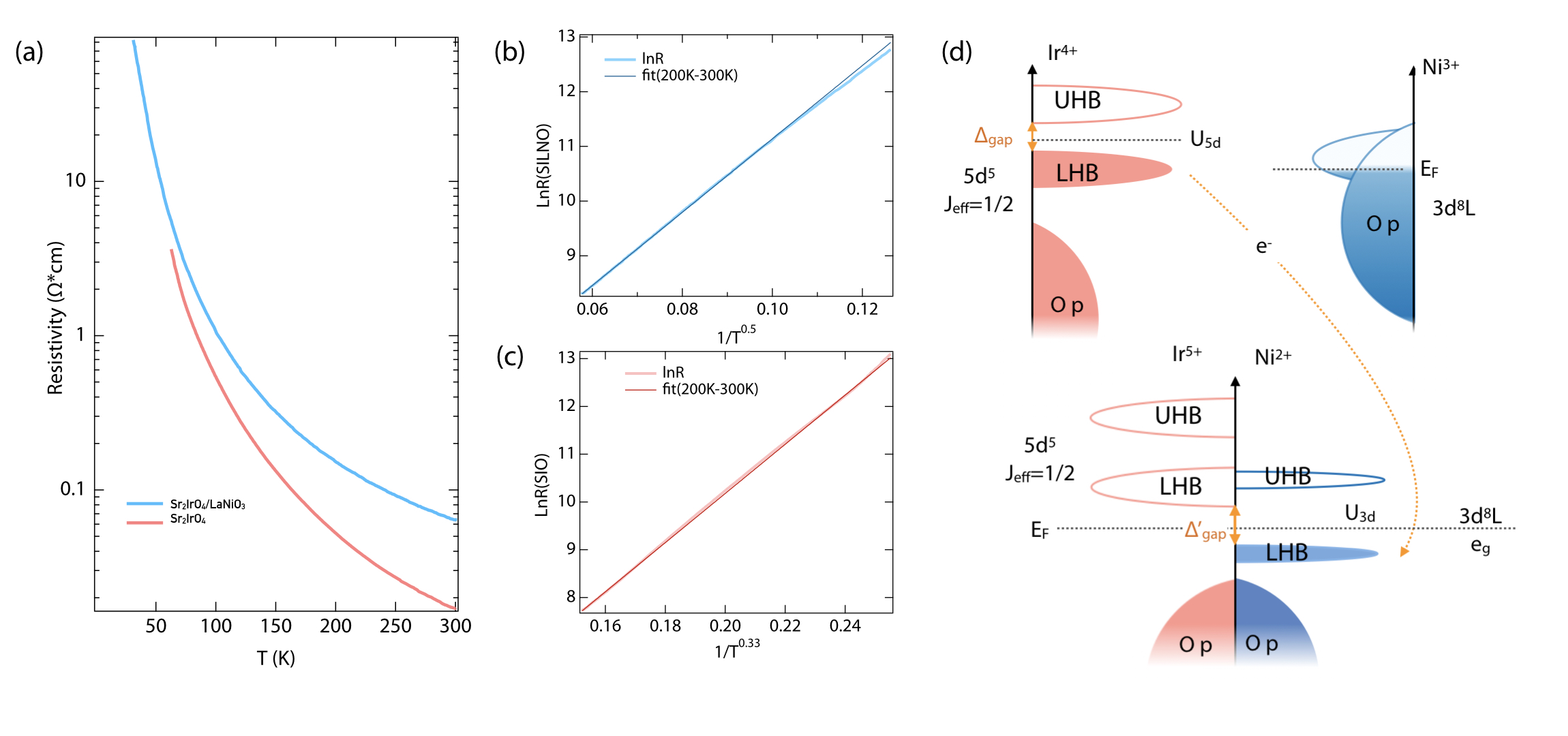}
    \caption{ 
    (a) Temperature-dependent resistivity of Sr$_2$IrO$_4$/LaNiO$_3$  and Sr$_2$IrO$_4$ single layer on the same SrTiO$_3$ substrate down to 2 K. Data was collected using the Van der Pauw method in a physical property measurement system. 
    (b) LnR vs 1/T$^{\frac{1}{2}}$ fitting result for Sr$_2$IrO$_4$/LaNiO$_3$.
    (c) LnR vs 1/T$^{\frac{1}{3}}$ fitting result for Sr$_2$IrO$_4$.
    (d) Schematic illustration of electronic reconstruction at Sr$_2$IrO$_4$/LaNiO$_3$ interface. 
    }
    \label{RT}
\end{figure*}

After establishing the orbital reconstructions at the interfaces, a critical question yet remains: Has the hole doping of Sr$_2$IrO$_4$ lead to the formation of a metallic phase? 
To investigate this issue, the temperature dependence of resistivity on both the superlattice and undoped Sr$_2$IrO$_4$ thin film were carried out from 300 K down to 2 K. 
As shown in Fig. \ref{RT}(a), an overall insulating behavior still persists in the superlattice, with the magnitude of resistivity exceeding the measurable limit at $\sim$ 30 K. 
By fitting the transport result with different variable range hopping (VRH) models:  $\sigma=\sigma_0exp\Big\{\Big(\frac{T_0}{T}\Big)^{1/(\beta+1)}\Big\}$ under all different $\beta$ selection ($\beta=0,1,2,3$) with the same temperature range, we found out that the best fitting model for the doped and undoped system are different. In undoped Sr$_2$IrO$_4$, it fits well with $\beta=2$, indicating a conventional 2D Mott variable range hopping(VRH) model (Fig. \ref{RT} (c)). This has also been observed in the undoped Sr$_2$IrO$_4$ from other groups, assuming that the band gap in the system is smaller than thermal activation ($\Delta_{gap}<k_B T$).\cite{LM2015Tuning,CL2014Crossover,Mott1969Conduction}
On the other hand, for doped Sr$_2$IrO$_4$, the best fitting result comes from the $\beta=1$ case, which is the Efros-Shklovskii VRH model (Fig. \ref{RT} (b)). In this case, a Coulomb gap larger than the thermal activation ($\Delta_{gap}>k_B T$) is opened, and the electrons becomes more localized after hole doping.\cite{ES1975Coulomb} 
In the following, we propose a possible scenario to interpret the electronic reconstruction at Sr$_2$IrO$_4$/LaNiO$_3$ interface.      

As seen in Fig. \ref{RT} (d), bulk LaNiO$_3$  is a negative charge-transfer metal, whose Fermi surface lies in hybridized O 2$p$ and Ni 3$d$ state\cite{dobin2003electronic,guo2018antiferromagnetic}.
Due to the strong O-Ni hybridization, the electronic structure of bulk LaNiO$_3$ is best  represented as a mixture of  3d$^7$ , 3d$^8$\underline{L} and 3d$^9$ \underline{LL} (\underline{L} is a hole on oxygen 2p orbital).  \cite{Golalikhani2018nature}
For Sr$_2$IrO$_4$, the Fermi surface is between the LHB and UHB of the J$_\textrm{eff}$=1/2 band. 
Driven by  the chemical potential difference across the Sr$_2$IrO$_4$/LaNiO$_3$ interfaces, electrons are transferred from the Ir to the Ni site.
Ir$^{4+}$ becomes Ir$^{5+}$ with the fully occupied J$_\textrm{eff}$ = 3/2 band still below the Fermi level.
The electron then can fill the hole on oxygens, and reduce the degree of  hybridization by lowering the oxygen p orbitals and split the Ni 3d band due to the on-site Coulomb repulsion, U. 
Since on Ni U is rather large ($\sim$ 4 eV), splitting of the Ni e$_g$ band will likely push the Ni up Hubbard band (UHB)  away from the Fermi level and above the Ir J$_\textrm{eff}$ = 1/2 low Hubbard band (LHB) bands. 
As the result, the band gap at the interface is determined by the  energy interval between the top of Ni e$_g$ LHB and the bottom of Ir J$_\textrm{eff}$ = 1/2 LHB.    

\section{CONCLUSION}

In conclusion, by growing a Sr$_2$IrO$_4$/LaNiO$_3$ superlattice, we have achieved hole doping of Sr$_2$IrO$_4$  without detrimental  effects of chemical disorder. 
XPS and XAS measurements confirm a charge transfer from Ir$^{4+}$ to Ni$^{3+}$, resulting in a markedly  fraction of Ir$^{5+}$ and Ni$^{2+}$. The latter is accompanied by a lifting of degeneracy in the Ni $e_g$ orbitals, with the d$_{3z^2-r^2}$ orbital ~0.1 eV lower than d$_{x^2-y^2}$, as evidenced by XLD measurements. Hole doping via the interface enhanced the on-site Coulomb interaction and enlarged the magnitude of the band gap in the Sr$_2$IrO$_4$/LaNiO$_3$  heterostructure compared to a Sr$_2$IrO$_4$ single layer. These results explore the phase diagram of Sr$_2$IrO$_4$ on the hole-doping side, and pave an alternative way towards the modification of carrier concentration in the relativistic Mott insulator Sr$_2$IrO$_4$.
 
 \section{Acknowledgements}
 
F. W., X. L. and J. C. acknowledged the support by the Gordon and Betty Moore Foundation EPiQS Initiative through Grant No. GBMF4534. M. K. and B. P. were supported by the Department of Energy Grant No. DE-SC0012375.  Q. Z. and L. G. were supported by the Strategic Priority Research Program of Chinese Academy of Sciences (Grant No. XDB07030200) and National Natural Science Foundation of China (51522212, 51421002, and 51672307).  This research used resources of the Advanced Light Source, which is a DOE Office of Science User Facility under contract no. DE-AC02-05CH11231.

\end{document}